\documentclass[%
 reprint,
superscriptaddress,
 amsmath,amssymb,
 aps,
prb,
]{revtex4-2}

\setcitestyle{super,sort&compress}

\usepackage{microtype}
\usepackage{amsmath}
\usepackage{graphicx}
\usepackage{dcolumn}
\usepackage{bm}
\usepackage{adjustbox} 
\usepackage{graphicx}

\usepackage{color, hyperref}
\definecolor{mygrey}{gray}{0.80}
\definecolor{darkblue}{RGB}{8,81,156}
\hypersetup{colorlinks,breaklinks,
            linkcolor=darkblue,urlcolor=darkblue,
            anchorcolor=darkblue,citecolor=darkblue}

\definecolor{super-dark-green}{RGB}{0,69,41}
\definecolor{super-dark-purple}{RGB}{63,0,125}
\definecolor{super-dark-blue}{RGB}{8,48,107}
\definecolor{super-dark-red}{RGB}{165,0,38}

\begin{document}


\title{Robustness of classical nucleation theory to chemical heterogeneity of crystal nucleating substrates}

\author{Fernanda Sulantay Vargas}
 \email{fernanda.vargas@yale.edu}
\author{Sarwar Hussain}%
\author{Amir Haji-Akbari}
 \email{amir.hajiakbaribalou@yale.edu}
\affiliation{%
 Department of Chemical and Environmental Engineering,\\
 Yale University, New Haven, Connecticut 06520, United States
}%
\date{\today}

\begin{abstract}
\noindent
{\bf Abstrtact:}
Heterogeneous nucleation is a process wherein extrinsic impurities facilitate freezing by lowering nucleation barriers and constitutes the dominant mechanism for crystallization in most systems. Classical nucleation theory (\textsc{Cnt}) has been remarkably successful in predicting the kinetics of heterogeneous nucleation, even on chemically and topographically non-uniform surfaces, despite its reliance on several restrictive assumptions, such as the idealized spherical-cap geometry of the crystalline nuclei. Here, we employ molecular dynamics simulations and jumpy forward flux sampling to investigate the kinetics and mechanism of heterogeneous crystal nucleation in a model atomic liquid. We examine both a chemically uniform, weakly attractive liquiphilic surface and a checkerboard surface comprised of alternating liquiphilic and liquiphobic patches. We find the nucleation rate to retain its canonical temperature dependence predicted by \textsc{Cnt} in both systems. Moreover, the contact angles of crystalline nuclei exhibit negligible dependence on nucleus size and temperature. On the checkerboard surface, nuclei maintain a fixed contact angle through  pinning at patch boundaries and vertical growth into the bulk. These findings offer insights into the robustness of \textsc{Cnt} in experimental scenarios, where nucleating surfaces often feature active hotspots surrounded by inert or liquiphobic domains.
\end{abstract}

\maketitle


\section{Introduction}

\noindent

\noindent
Crystallization is a first-order phase transition that underpins many important natural and industrial processes, from atmospheric ice nucleation\cite{Baker1997} and biomineralization\cite{VeisCalcTissueInt2013} to manufacturing of nanoparticles,\cite{BassaniACSNano2024} solar cells,\cite{SharmaNanoscaleAdv2021} semiconductors\cite{LeeNatRevMater2016} and pharmaceuticals.\cite{Chen2011} It proceeds through a nucleation and growth mechanism,\cite{Oxtoby1992HomogeneousExperiment} wherein thermal fluctuations within the metastable liquid lead to the formation of a sufficiently large crystalline nucleus, which then grows until the system reaches thermodynamic equilibrium. While crystals can nucleate homogeneously within a supercooled liquid or supersaturated solution, they often nucleate heterogeneously at extrinsic interfaces (e.g.,~provided by an impurity). Such impurities enhance nucleation kinetics by decreasing nucleation barriers, and manipulating their chemistry and topography is a feasible route for controlling and understanding the crystallization of proteins,\cite{Artusio2023Self-AssembledNucleation} minerals, \cite{Aizenberg1999ControlMonolayers, Guan2025GypsumHydrophobicity} pharmaceuticals\cite{Cox2007SelectivePolymorphs, Nordquist2025GrowthOn} and water.\cite{Varanasi2009SpatialWater}

The most widely used theoretical framework for describing crystal nucleation is the classical nucleation theory (\textsc{Cnt}).\cite{KarthikaCrystGrowthDes2016} While originally developed\cite{Volmer1926KeimbildungGebilden} for homogeneous vapor condensation, \textsc{Cnt} was later generalized to homogeneous nucleation within condensed phases.\cite{Becker1935KinetischeDampfen, Turnbull1949RateSystems} \textsc{Cnt} assumes crystalline nuclei that are spherically shaped and that possess sharp liquid-solid interfaces. The free energy of formation of a nucleus of radius $r$ is estimated as $\Delta{G}_f(r) = -\frac43\pi r^3|\Delta\mu| + 4\pi r^2\gamma_{ls}$ with $|\Delta\mu|$ and $\gamma_{ls}$ the thermodynamic driving force for crystallization and the liquid-solid surface tension, respectively.  The nucleation rate is therefore given by,
\begin{eqnarray}
R_{\text{hom}} &=& A_{\text{hom}}\exp\left[
-\frac{\Delta G^*_{\text{hom}}}{kT}
\right].\label{eq:rate-CNT}
\end{eqnarray}
Here, $A_{\text{hom}}$ is a kinetic prefactor that, among other things, primarily depends on diffusivity within the supercooled liquid, while $\Delta G^*_{\text{hom}}$ is the nucleation barrier that constitutes the maximum of $\Delta{G}_f$ and is given by,
\begin{eqnarray}
\Delta{G}^*_{\text{hom}} &=& \frac{16\gamma_{ls}^3}{3\rho_s^2|\Delta\mu|^2},
\end{eqnarray}
with $\rho_s$, the molar density of the crystal. An extension of \textsc{Cnt} to heterogeneous nucleation was devised by Turnbull,\cite{TurnbullJCP1950} wherein crystalline nuclei are assumed to be spherical caps that always maintain a fixed contact angle, $\theta_c$, at the crystal nucleating substrate.  In this formalism, rate is expressed in a manner similar to Eq.~\eqref{eq:rate-CNT} with a scaled nucleation barrier given by,
\begin{eqnarray}\label{eq:DelG-het}
\Delta{G}_{\text{het}}^* &=& f_c(\theta_c)\Delta{G}^*_{\text{hom}}.
\end{eqnarray}
The scaling factor $f_c(\theta_c)$ is often called  the potency factor\cite{Cabriolu2015IceNucleation}-- or compatibility factor\cite{AlpertPCCP2011}-- and is given by:
\begin{eqnarray}
f_c(\theta_c) &=& \tfrac14\left(1-\cos\theta_c\right)^2\left(2+\cos\theta_c\right).
\end{eqnarray}
The potency factor bears a clear geometric interpretation, corresponding to the relative size of the critical nucleus in heterogeneous vs. homogeneous nucleation. 

Despite its simplicity, \textsc{Cnt} has been remarkably successful in interpreting heterogeneous nucleation experiments.\cite{KarlssonBiophysJ1993, ConradJCP2005, WangMacromolecules2019} This success raises an interesting paradox, as \textsc{Cnt} assumes that crystalline nuclei maintain a fixed contact angle-- a condition strictly valid only on pristine, chemically and topographically uniform surfaces. In contrast, most real nucleating surfaces are inherently heterogeneous and dynamic, and possess defects and grain boundaries. Even on uniform surfaces, computational studies suggest that \textsc{Cnt} could fail in the fully wetting limit,\cite{AuerPRL2003} necessitating accounting for line tension contributions. While deviations from \textsc{Cnt} are reported in the literature,\cite{QiuJACS2019, Hussain2021RoleWater} the extent to which it remains valid in the presence of surface heterogeneity has yet to be systematically explored.

Here, we use molecular dynamics (\textsc{Md}) simulations and the jumpy forward flux sampling\cite{Haji-Akbari2018Forward-fluxParameters} (\textsc{jFfs}) algorithm to probe the kinetics of heterogeneous crystal nucleation of the Lennard-Jones\cite{LennardJonesPRoySocA1924} (\textsc{Lj}) liquid on patterned surfaces. In particular, we consider checkerboard surfaces with attractive and repulsive patches, and examine the dependence of nucleation rate on temperature to demonstrate the surprising robustness of \textsc{Cnt} to surface heterogeneity. We also probe the evolution of the microscopic contact angle during the nucleation process, observing a pinning mechanism that maintains an almost fixed contact angle for sufficiently large nuclei.

\section{Methods}

\subsection{Molecular Dynamics Simulations}\label{md_simulations}

\begin{figure}
    \centering
    \includegraphics[width=.5\textwidth]{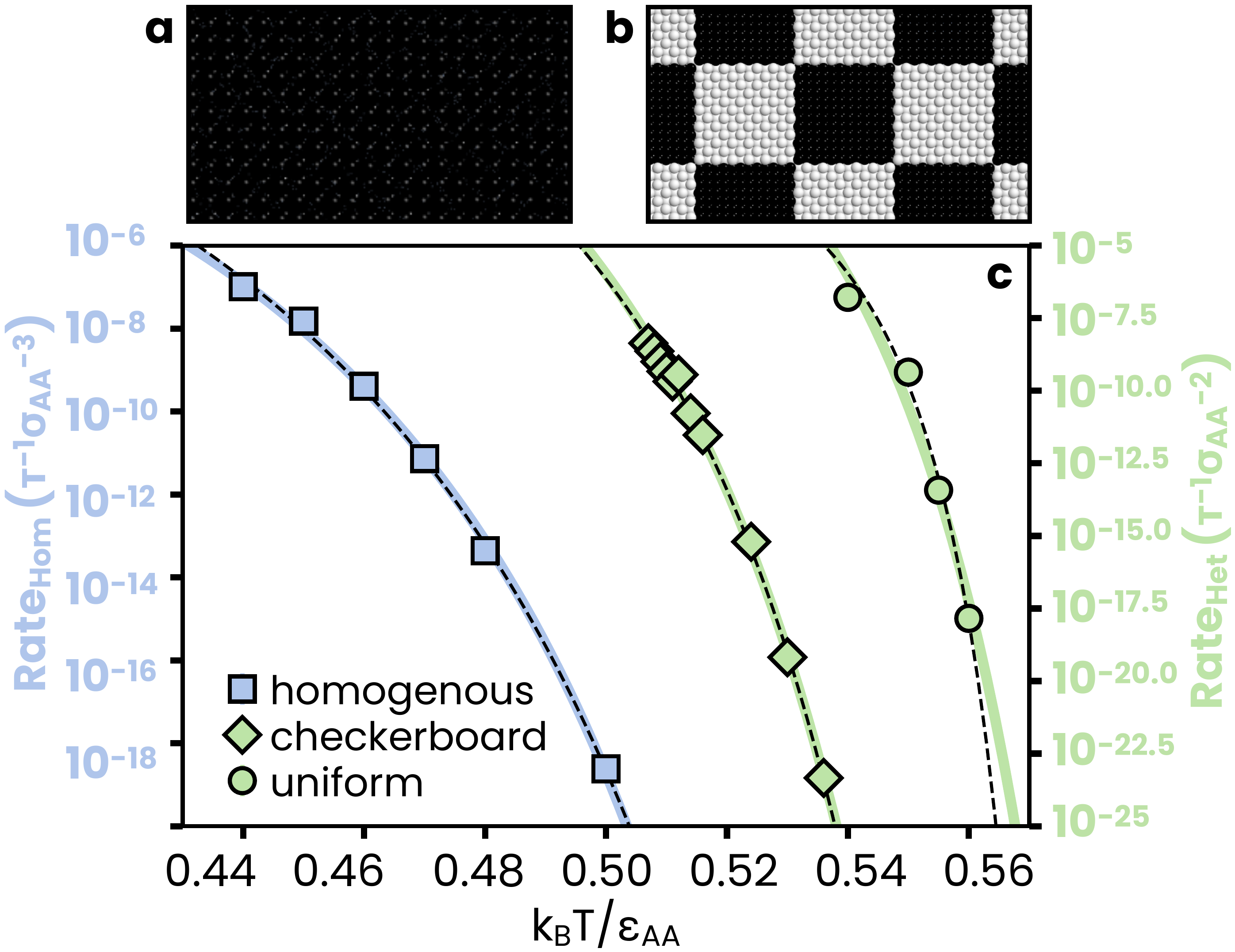}
    \caption{\label{fig:CNT_fits} Schematic representations of (a) a chemically uniform and (b) a checkerboard-patterned surface. Black and light gray particles correspond to liquiphilic and liquiphobic patches, respectively. (c) Temperature dependence of homogeneous and heterogeneous nucleation rates in the \textsc{Lj} system. The symbols correspond to actual rates computed from jFFS with error bars smaller than symbol sizes. The curves correspond to \textsc{Cnt} fits according to Eqs.~\eqref{eq:CNT-T-hom} and \eqref{eq:CNT-T-het} with $T_{m}$ as a fixed parameter (solid line) and as a fitting parameter (dashed line). Homogeneous nucleation rates are obtained from Refs.~\citenum{Haji-Akbari2018Forward-fluxParameters} and \citenum{HussainJCP2022}.}
\end{figure}

\noindent
We perform \textsc{Md} simulations using \textsc{Lammps},\cite{Thompson2022LAMMPSScales} integrating Newton’s equation of motion with the velocity Verlet algorithm and a reduced time step of $\delta{t}^*=2.5\times10^{-3}$ where $t^{*} = (\epsilon_{AA}/m_{A}\sigma_{AA}^{2})^{1/2}t$. All particles interact via a truncated and shifted \textsc{Lj} potential,\cite{LennardJonesPRoySocA1924} with interaction parameters summarized in Table~\ref{table:LJparameters}. In addition to the A particles that comprise the supercooled liquid, the system includes two types of wall particles: B and C. Liquiphilic patches (shown in black in Figs.~\ref{fig:CNT_fits}a-b) are composed of B particles, which interact with A particles through a weakly attractive \textsc{Lj} potential. In contrast, liquiphobic patches (depicted in light gray in Fig.~\ref{fig:CNT_fits}b) contain  C particles, which interact with A particles via the purely repulsive Weeks-Chandler-Andersen (\textsc{Wca}) potential.\cite{WeeksJChemPhys1971}

\begin{table}
        \caption{Interaction parameters for the \textsc{Lj} potential. Note that BB, BC, and CC interactions are turned off since the corresponding atoms are already bonded to one another in the corresponding substrates. }
        \label{table:LJparameters}
        \begin{tabular}{c c c c c} 
        \hline\hline
         ~~~$i$~~~ & ~~~$j$~~~  & ~~~$\epsilon_{ij}$~~~ & ~~~$\sigma_{ij}$~~~ & ~~~$r_{c,ij}$~~~ \\ [0.5ex] 
         \hline
         ~~~A & ~~~A & ~~~1.0 & ~~~1.0 & ~~~2.5 \\ 
         ~~~A & ~~~B & ~~~0.5 & ~~~1.0 & ~~~2.5 \\
         ~~~A & ~~~C & ~~~0.3 & ~~~1.0 & ~~~1.1225 \\ [1ex] 
         \hline
        \end{tabular}
\end{table}

All simulation boxes are periodic along all dimensions and are comprised of the supercooled liquid confined within a slit pore formed by a nucleating substrate and a repulsive wall (composed of particles of type C). This setup, which is distinct from the mirroring approach\cite{SossoJChemPhys2016, BiJPhysChemC2016} employed in some earlier studies, allows us to probe the exclusive effect of the isolated crystal nucleating surface without inducing spurious nano-confinement effects.  We consider two types of nucleating substrates: a uniform surface comprised of particles of type B only, and a checkerboard surface comprised of almost-square-shaped liquiphilic and liquiphobic patches of side length $\approx9\sigma_{AA}$. Both walls are face-centered cubic (\textsc{Fcc}) lattices of reduced number density $\rho^*_n=\rho_n\sigma_{AA}^3 =1.13$, with their 001 planes exposed to the liquid. Interactions among wall particles are turned off; instead, neighboring wall particles are connected via harmonic springs with an equilibrium bond length of $r_0 = 1.08\,\sigma_{AA}$ and a spring constant of $k = 500\,\epsilon_{AA}/\sigma_{AA}^2$. Table~\ref{tab:system-sizes} summarizes further details about each system, including the number of particles of each type, as well as the average dimensions of the simulation boxes.

\begin{table*}
\centering
\caption{Number of particles of each type, as well as the average dimensions of the simulation box in each system. $n_s$ and $n_r$ refer to the number of particles within the nucleating substrate and the repulsive wall, respectively. Note that we use a larger system size to probe nucleation on the chemically uniform surface at higher temperatures to avoid finite-size effects.  \label{tab:system-sizes}}
\begin{tabular}{l|c|ccc|cc|ccc}
\hline\hline
System~~~~~~~~ & ~~$kT/\epsilon_{AA}$~~ &~~$n_A$~~ & ~~$n_B$~~ & ~~$n_C$~~ & ~~$n_s$~~ & ~~$n_r$~~ & ~~$l_x\,[\sigma_{AA}]$~~ & ~~$l_y\,[\sigma_{AA}]$~~ & ~~$l_z\,[\sigma_{AA}]$~~ \\
\hline
Uniform & 0.54, 0.55 & 10,752 & 2,048 & 2,048 & 2,048 & 2,048 & 24.38 & 24.38 & $\approx29$\\
Uniform & 0.555, 0.56 & 24,192 & 4,608 & 4,608 & 4,608 & 4,608 & 36.57 & 36.57 & $\approx29$\\
Checkerboard & All & 24,192 & 2,304 & 6,912 & 4,608 & 4,608 & 36.57 & 36.57 & $\approx28$ \\
\hline
\end{tabular}
\end{table*}

All simulations are conducted in the isothermal-isotension ($Np_zT$) ensemble  where pressure is only controlled along the $z$ direction, while the $x$ and $y$ dimensions of the simulation box are kept fixed. Temperature and pressure are controlled using the Nos\'{e}-Hoover thermostat\cite{NoseMolPhys1984, HooverPhysRevA1985} and the Parrinello-Rahman barostat,\cite{ParrinelloJAppPhys1981} with coupling time constants of $10^2\delta{t}$ and $10^3\delta{t}$, respectively.

All simulations are initiated from a configuration in which liquid and wall particles are arranged on an \textsc{Fcc} lattice of reduced number density $\rho_n^*=1.13$. Bonds are introduced between neighboring wall particles-- defined as those separated by a distance between $1.07\sigma_{AA}$ and $1.09\sigma_{AA}$. The system is then melted at a temperature of $kT/\epsilon_{AA} = 1$ and a pressure of $p\sigma_{AA}^3/\epsilon_{AA} = 1$. Following an initial $5 \times 10^4$-step equilibration period, during which the crystal within the slit pore fully melts, 100--200 configurations are saved every $10^4$ steps. These configurations serve as starting points for quenching \textsc{Md} trajectories in which the temperature and pressure are gradually reduced to $kT/\epsilon_{AA} = 0.75$ and $p\sigma_{AA}^3/\epsilon_{AA} = 0.05$ over $2.5 \times 10^4$ steps. Subsequently, the arising configurations are further quenched to the final target temperature at constant pressure. The resulting configurations are then briefly equilibrated at the target state point and used as starting configurations for basin exploration in the \textsc{jFfs} calculations.

\subsection{Rate Calculations using jumpy forward-flux sampling}

\noindent
 We compute the nucleation rates by using the \textsc{jFfs} algorithm,\cite{Haji-Akbari2018Forward-fluxParameters} an extension of conventional forward-flux sampling\cite{AllenJCP2006} designed to accommodate order parameters that exhibit discontinuous temporal jumps. Such jumps are unavoidable for order parameters describing aggregation-driven processes such as nucleation. The order parameter employed here is  the size of the largest crystalline nucleus, determined using neighbor-averaged\cite{Lechner2008AccurateParameters} Steinhardt bond order parameters,\cite{Steinhardt1983} augmented by the pruning procedure outlined in Ref.~\citenum{Domingues2024DivergenceCrystals}. Employing pruning is key to assuring the efficacy of the employed order parameter, which has been shown to be strongly impacted by implementation details.\cite{Domingues2024DivergenceCrystals, SinaeianJCP2025}

 To sample the initial supercooled liquid basin, we perform a minimum of 100 \textsc{Md} trajectories initiated from independent configurations, with a total simulation time exceeding $5 \times 10^5\,\sigma_{AA}/\sqrt{m_A\epsilon_{AA}}$. Each subsequent \textsc{jFfs} iteration is terminated after recording at least 3,000 successful crossings. The computed rates are given in Tables~\ref{table:LJ_hetrates_good} and \ref{table:LJ_hetrates_chck}.  System sizes are chosen conservatively to eliminate finite-size artifacts.\cite{Hussain2021HowNucleation} In particular, a larger system size is utilized for the uniform surface at $kT/\epsilon_{AA}=0.555$ and $0.56$, as our initial rate calculations exhibited strong finite size effects.

We need to emphasize that the use of the \textsc{jFfs} algorithm is absolutely necessary for computing the nucleation rates that vary by as much as 15 orders of magnitude. Even at lower temperatures where spontaneous nucleation is occasionally observed during conventional \textsc{Md}, the rates computed from \textsc{jFffs} are statistically indistinguishable from those estimated using the mean first passage time\cite{WedekindJCP2007} (\textsc{Mfpt}) method. For methodological consistency, all the rates reported here were computed using \textsc{jFfs} calculations.

\begin{table}[h]
    \centering
    \caption{Heterogeneous nucleation rates in the uniform surface at $p^{*} = 0.05$. Reported uncertainties correspond to 95\% confidence intervals.}
    \label{table:LJ_hetrates_good}
    \begin{tabular}{c|c} 
        \hline\hline
        $kT/\epsilon_{AA}$  & $\log_{10}R_{\text{het}} \left[\sigma_{AA}^{-3}\epsilon_{AA}^{-1/2}m_A^{-1/2}\right]$\\ [0.5ex] 
        \hline
        $0.540$ & ~$-6.7813 \pm 0.0384$ \\ 
        $0.550$ & ~$-9.3574 \pm 0.0509$ \\
        $0.555$ & $-13.4224 \pm 0.0939$ \\
        $0.560$ & $-17.8410 \pm 0.1119$ \\ [1ex] 
        \hline
        \hline
    \end{tabular}
\end{table}

\begin{table}[h]
    \centering
    \caption{Heterogeneous nucleation rates in the checkerboard system at $p^*=0.05$. Reported uncertainties correspond to 95\% confidence intervals.}
    \label{table:LJ_hetrates_chck}
    \begin{tabular}{c|c} 
        \hline\hline
        $kT/\epsilon_{AA}$  & $\log_{10}R_{\text{het}} \left[\sigma_{AA}^{-3}\epsilon_{AA}^{-1/2}m_A^{-1/2}\right]$\\ [0.5ex] 
        \hline
        0.507 & ~$-8.3630 \pm 0.0487$ \\ 
        0.508 & ~$-8.6354 \pm 0.0502$ \\
        0.509 & ~$-8.9985 \pm 0.0513$\\
        0.510 & ~$-9.3320 \pm 0.0549$\\ 
        0.511 & ~$-9.6795 \pm 0.0554$\\
        0.512 & ~$-9.4458 \pm 0.0648$\\
        0.514 & $-10.7786 \pm 0.0634$\\ 
        0.516 & $-11.5264 \pm 0.0705$\\
        0.524 & $-15.2021 \pm 0.0900$\\
        0.530 & $-19.1809 \pm 0.1090$\\
        0.536 & $-23.3385 \pm 0.1248$\\ [1ex] 
        \hline\hline
    \end{tabular}
\end{table}

\subsection{Contact Angle Calculations}
\label{section:contact-angle-calc}

\begin{figure}[h]
    \centering
    \includegraphics[width=.3\textwidth]{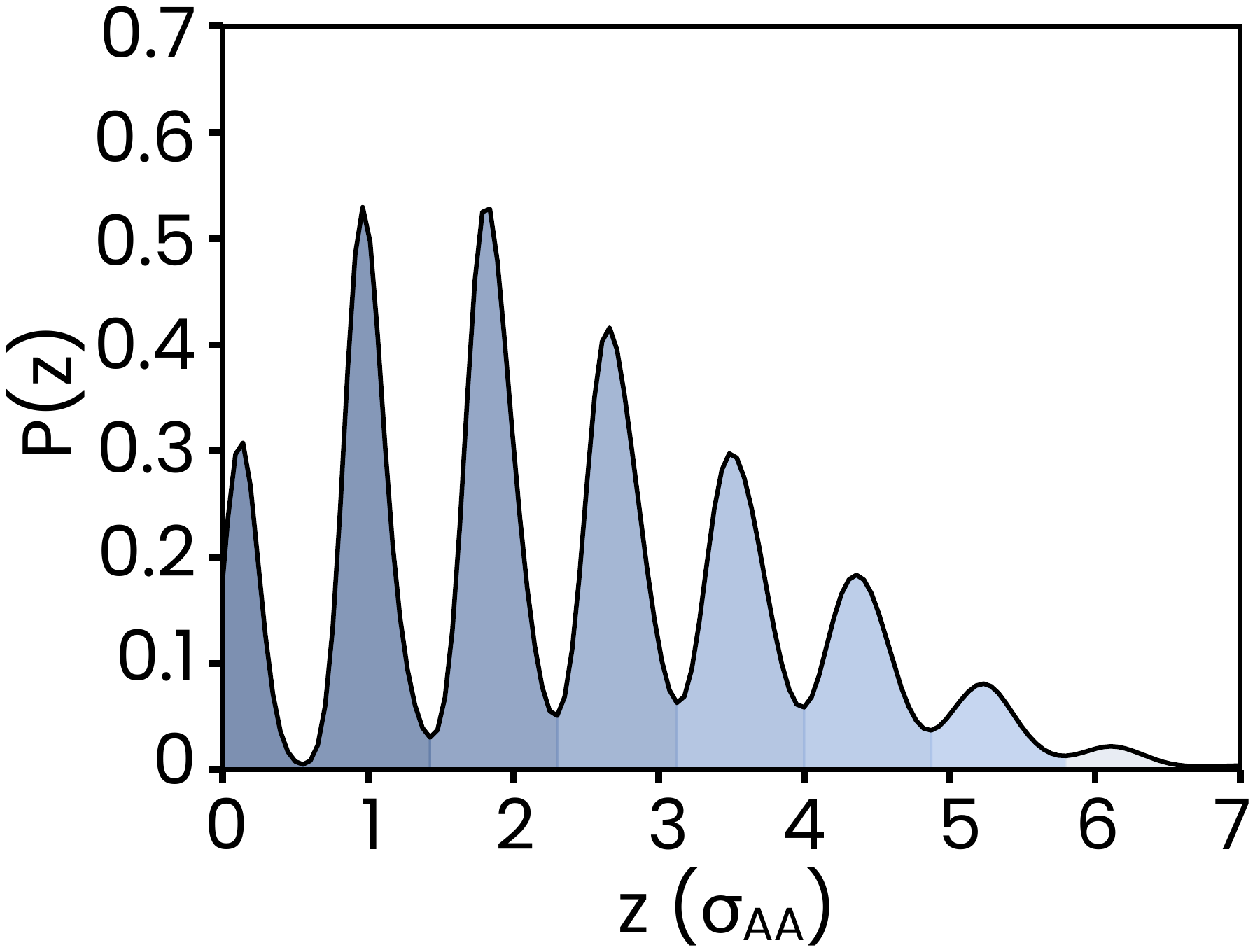}
    \caption{\label{fig:cluster_densityprofile} $p(z)$ of crystalline nuclei recorded at the last \textsc{jFfs} milestone (corresponding to $\sim262$ particles) of the rate calculation conducted in the checkerboard system at $k_BT/\varepsilon_{AA}=0.507$. }
\end{figure}

\noindent
To estimate the microscopic contact angles of the crystalline nuclei, we follow the procedure outlined in Ref.~\citenum{Giovambattista2007EffectStructure}. For each temperature and system type, we analyze the configurations recorded at the last \textsc{jFfs} milestone to compute $p(z)$, the probability density that the $z$ coordinate of a given particle within the nucleus is at a normal distance $z$ from the bottom of that nucleus. Fig.~\ref{fig:cluster_densityprofile} illustrates one such $p(z)$ in the checkerboard system. It is important to note that $p(z)$  serves as a proxy for the number density profile within a typical nucleus,  which is challenging to systematically normalize due to the irregular shapes of individual nuclei.  The $p(z)$ profiles computed from configurations at earlier milestones exhibit a qualitatively similar pattern, albeit with fewer peaks.

\begin{figure*}
    \centering
    \includegraphics[width=.9\textwidth]{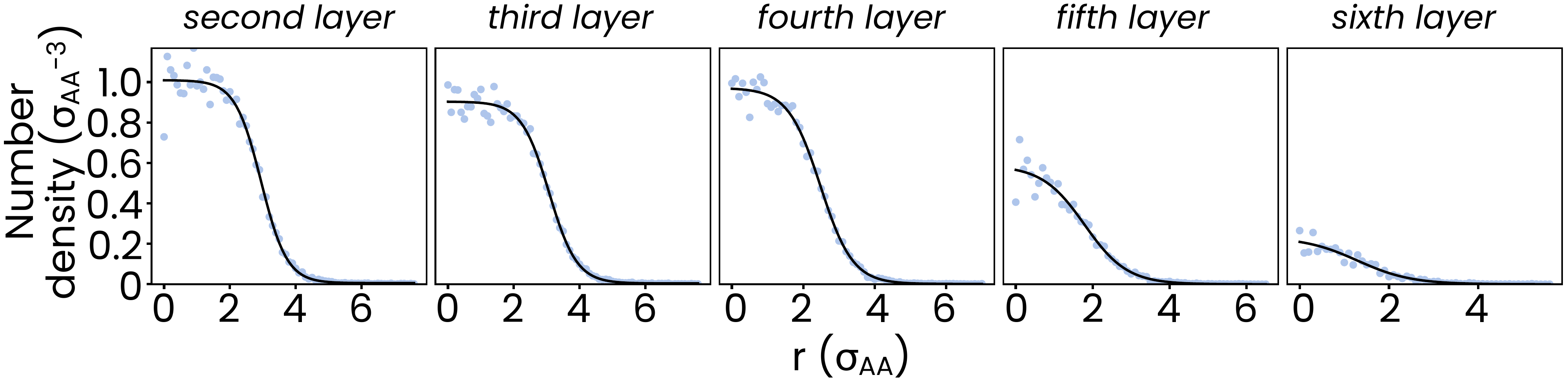}
    \caption{\label{fig:radial_densityprofile} Radial number density profiles alongside their respective hyperbolic tangent  fits for the 2nd-6th layers of crystalline nuclei comprised of $\sim91$ particles obtained at $k_BT/\varepsilon_{AA}=0.510$ in the checkerboard system. (First layer excluded due to peculiarities of pruning.)
    }
\end{figure*}

\begin{figure}
    \centering
    \includegraphics[width=.3\textwidth]{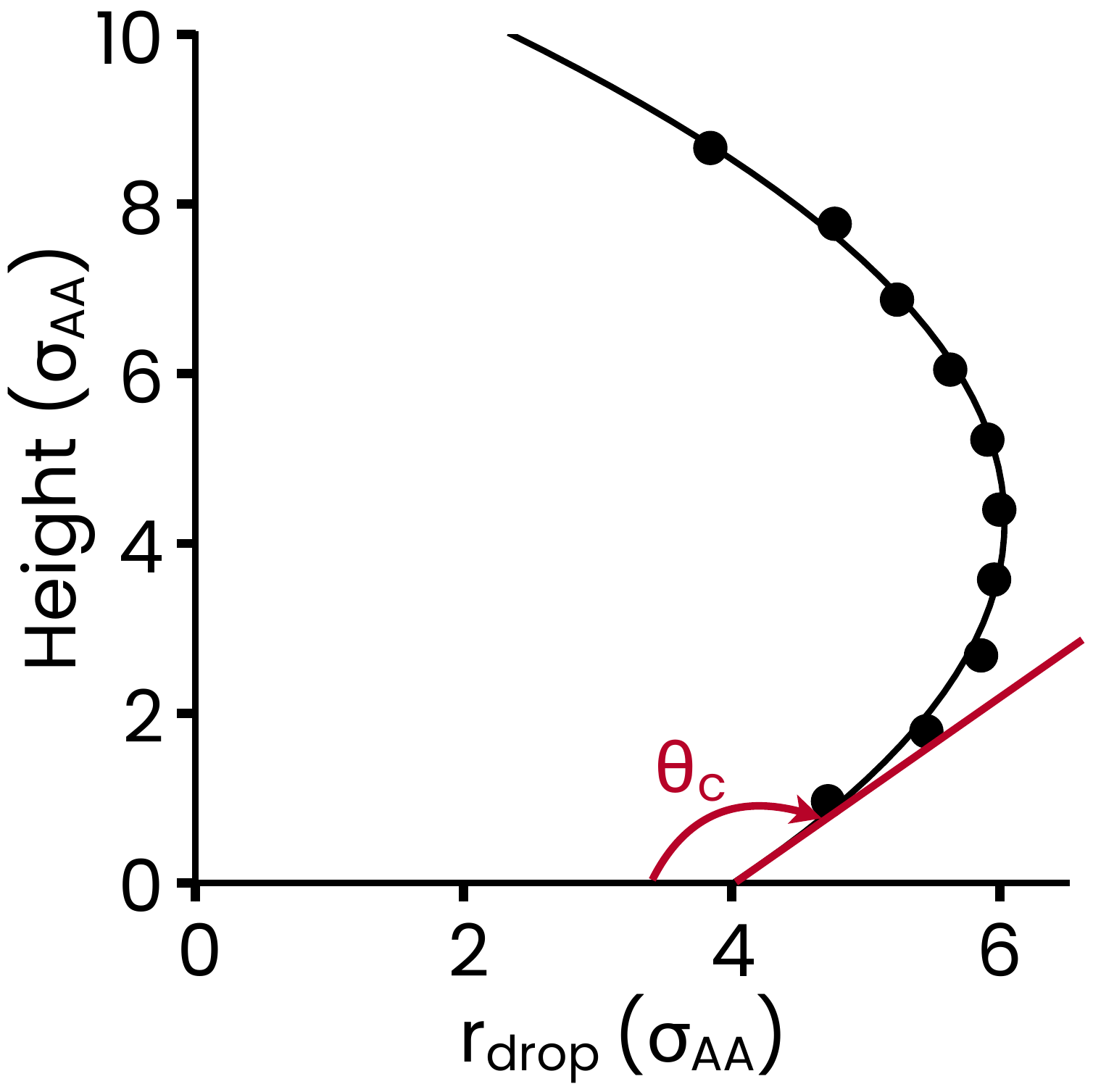}
    \caption{\label{fig:dropprofile} Drop profile of a crystalline nucleus on the checkerboard surface at $k_BT/\varepsilon_{AA}=0.536$ with a cluster size of $\sim$841 particles. Filled circles correspond to individual $r_{\text{drop}}$'s computed for different layers, while the black curve corresponds to the quadratic fit. The estimated contact angle is illustrated in red. }
\end{figure}

Crystalline nuclei gathered at all \textsc{jFfs} milestones are then partitioned into layers demarcated by the valleys of $p(z)$. Within each layer, we compute $\rho(r,z)$, the radial number density profile as a function of $r$, the distance to the layer's respective center of mass. This profile is fitted to a hyperbolic tangent function of the form:
$$
\rho(r,z) = a\tanh\left[b(r-r_{\text{drop}}(z))
\right]  + c,
$$
which yields $r_{\text{drop}}(z)$, the radial boundary of the nucleus in the corresponding layer. Several typical radial number density profiles are illustrated in Fig.~\ref{fig:radial_densityprofile}. We then identify the layers that have well-developed crystalline regions as those that possess a plateau value exceeding $0.60$. If three or more layers satisfy this plateau requirement, their  $r_{\text{drop}}$'s are fitted to a quadratic function:
$$
r_{\text{drop}}(z) = a_0 + a_1z + a_2z^2,
$$
The contact angle, $\theta_c$, is subsequently estimated from the slope of the quadratic fit at the substrate (i.e., zero height) using the relation:
$$
\theta_c = \frac\pi2 + \tan^{-1} a_1
$$
Fig.~\ref{fig:dropprofile} depicts a typical drop profile. We wish to note that we exclude the $r_{\text{drop}}$ obtained from the first layer due to the peculiarities of the pruning algorithm~\cite{Domingues2024DivergenceCrystals} utilized in our order parameter.

To estimate the uncertainties in the computed contact angles, configurations at any given milestone are randomly divided into five groups. For each group, we separately estimate $\rho(r,z)$, $r_{\text{drop}}(z)$, and $\theta_c$. The reported contact angles are the means of such estimates, with the associated uncertainty given by the standard error among estimates from different groups. Varying the number of groups or the random number seed used for partitioning does not lead to statistically significant changes in contact angle estimates.

\begin{figure}[h]
    \centering
    \includegraphics[width=.4\textwidth]{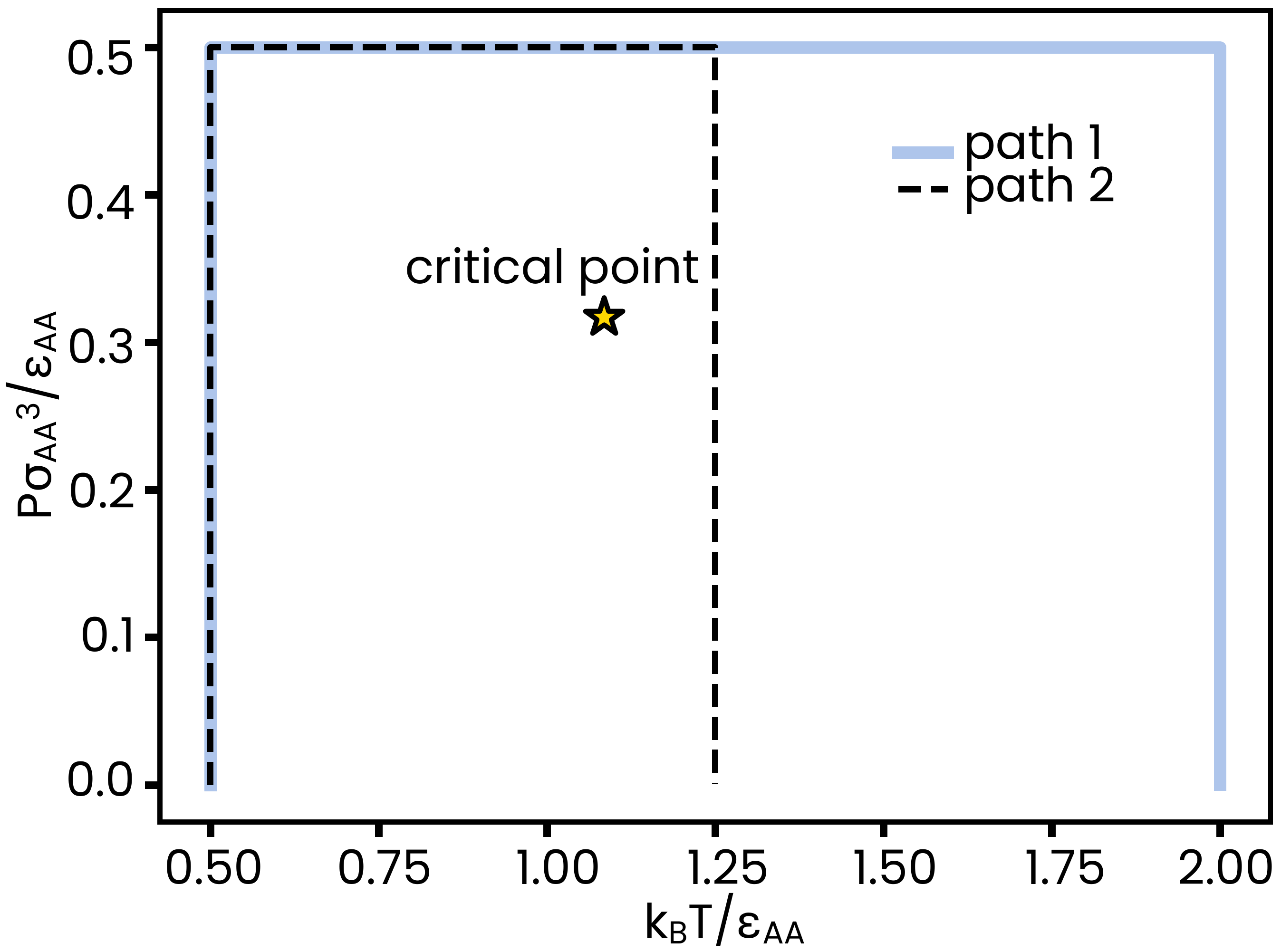}
    \caption{\label{fig:thermo_int_paths} Two distinct paths within the $p-T$ diagram utilized for the liquid thermodynamic integration in the \textsc{Lj} system. Both paths go around the critical point as reported in Ref.~\citenum{Smit1992} and yield identical chemical potentials for the liquid. }
\end{figure}

\begin{figure}
\centering
\includegraphics[width=.35\textwidth]{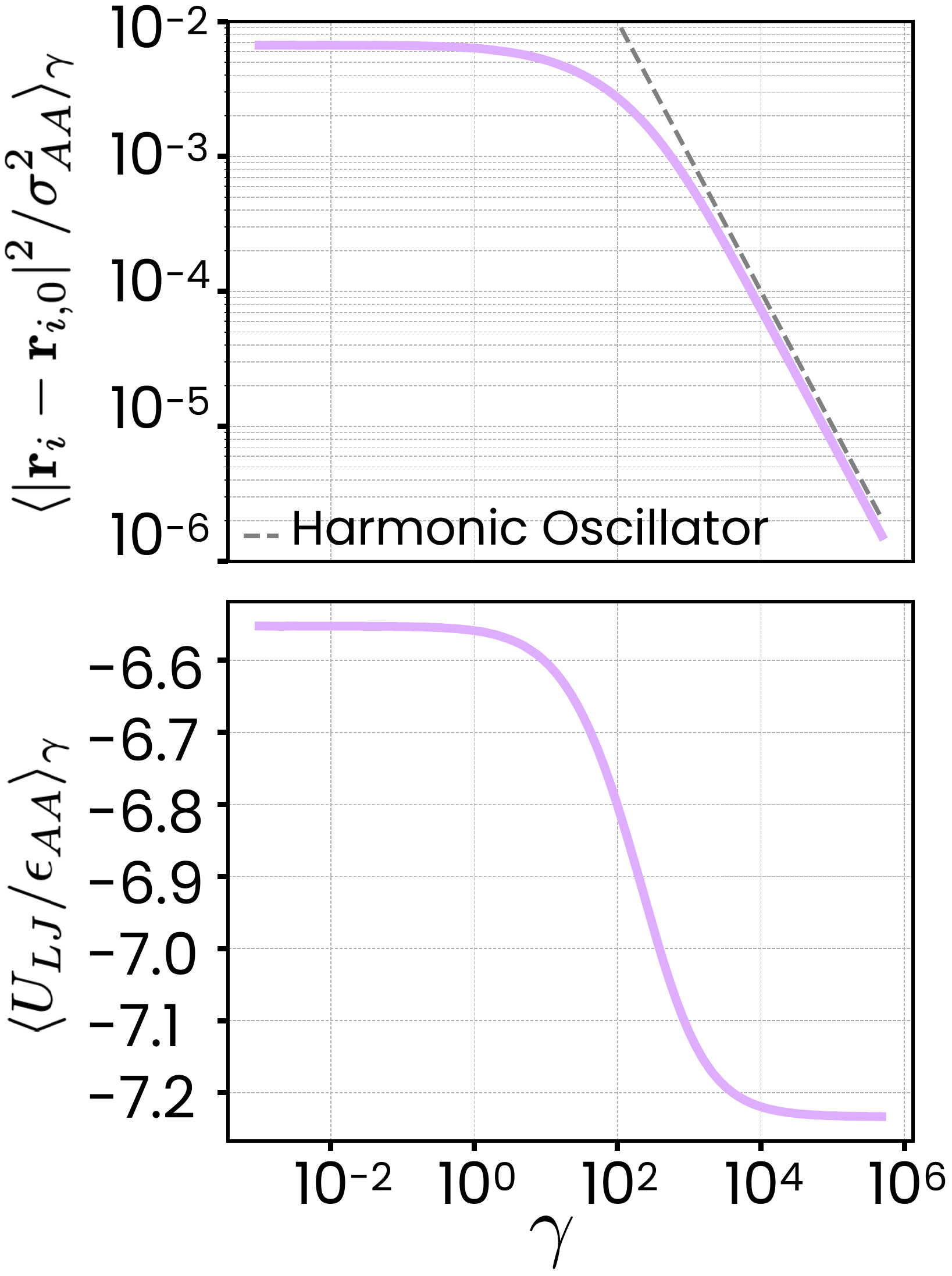}
\caption{(a) Mean-squared displacement and (b) average potential energy as a function of the coupling parameter, $\gamma$. The dashed line in (a) corresponds to the limiting behavior given by Eq.~\eqref{eq:limiting-harmonic-MSD}. \label{fig:Fcc-int}}
\end{figure}

\subsection{Free Energy Calculations}
\label{section:free-energy-calcs}
\noindent
We compute the free energy difference between the supercooled liquid and the \textsc{Fcc} crystal using thermodynamic integration combined with a variant of the Frenkel–Ladd method,\cite{FrenkelJCP1984} as originally described in Ref.~\citenum{Haji-Akbari2011PhaseTetrahedra}. To obtain the free energy of the supercooled liquid, we continuously transform it into an ideal gas along a sequence of isotherms and isobars, as illustrated in Fig.~\ref{fig:thermo_int_paths}. Note that doing so requires going around the liquid-vapor critical point.\cite{Smit1992} The configuration chemical potential of an ideal gas at temperature $T$ and pressure $p$ is given by,
$$
\mu_{\text{ideal}} (p,T;n) =  kT\ln\frac{p}{kT} + \frac{\ln 2\pi n}{n},
$$
where the second term accounts for the indistinguishability of particles and vanishes in the thermodynamic limit, $n\rightarrow\infty$, serving as a finite-size correction. Changes in chemical potential along an isotherm or an isobar are computed via:
\begin{eqnarray}
\mu(p_2,T) - \mu(p_1,T) &=& \int_{p_1}^{p_2} \left\langle v\right\rangle\,dp\label{eq:P-int}\\
\frac{\mu(p,T_2)}{T_2} - \frac{\mu(p,T_1)}{T_1} &=& -\int_{T_1}^{T_2} \frac{\langle h\rangle}{T^2}\,dT\label{eq:T-int}
\end{eqnarray}
where $\langle v\rangle$ and $\langle h\rangle$ correspond to average specific volume and average configurational enthalpy per particle, respectively. While these integrals can be evaluated numerically using the trapezoid rule, rapid changes in $\langle v\rangle$ and $\langle h\rangle$ might lead to considerable systematic errors. We therefore use an alternative numerical integration scheme that works as follows. For $\langle v\rangle$, we use the following functional form between two successive points $p_1$ and $p_2$ along an isobar:
\begin{eqnarray}
\langle v\rangle &\approx& \frac{f(p)}{p} = \frac{1}{p}\left[
v_1p_1+\frac{v_2p_2-v_1p_1}{p_2-p_1}\left(p-p_1\right)
\right]\notag\\
&=& \frac{(v_2-v_1)p_1p_2}{(p_2-p_1)p} + \frac{v_2p_2-v_1p_1}{p_2-p_1}.
\end{eqnarray}
This particular fitting procedure yields the following approximation of the integral:
\begin{eqnarray}
\int_{p_1}^{p_2}\langle v\rangle\, dp \approx  \frac{(v_2-v_1)p_1p_2}{p_2-p_1}\ln\frac{p_2}{p_1} + v_2p_2-v_1p_1,
\end{eqnarray}
which is more accurate and numerically more stable than the simple trapezoid estimate $\frac12(v_1+v_2)(p_2-p_1)$, particularly closer to the ideal gas limit. For an isotherm, a linear function is fit to two successive points, and the integral is estimated as:
\begin{eqnarray}
\int_{T_1}^{T_2}\frac{\langle h\rangle}{T^2}\,dT \approx \frac{h_1T_2-h_2T_1}{T_1T_2} + \frac{h_2-h_1}{T_2-T_1}\ln\frac{T_2}{T_1}.
\end{eqnarray}
These integrals are evaluated numerically along the specified thermodynamic paths in a disordered system comprised of 6,912 \textsc{Lj} particles.  The computed chemical potentials of the supercooled liquid are robust with respect to the choice of thermodynamic path and the discretization scheme used along each path for numerical integration.

To compute the free energy of the \textsc{Fcc} crystal, we continuously transform the crystal into an Einstein solid. This is accomplished by introducing a harmonic restraint potential, such that the system Hamiltonian becomes:
\begin{eqnarray}
\mathcal{H}\left(\textbf{r}^n;\gamma\right) &=& \mathcal{U}_{\text{LJ}}\left(\textbf{r}^n\right) + \frac{\gamma}{2\sigma_{AA}^2} \sum_{i=1}^n |\textbf{r}_i-\textbf{r}_{i,0}|^2,
\end{eqnarray}
where $n$ is the number of particles, $\mathcal{U}_{\textsc{LJ}}$  is the potential energy of the unperturbed system (according to the \textsc{Lj} potential),  $\textbf{r}^n \equiv (\textbf{r}_1,\textbf{r}_2,\cdots,\textbf{r}_n)$ are the instantaneous particle positions, and $\textbf{r}_0^n$ are the corresponding ideal lattice positions. For sufficiently large values of $\gamma$, fluctuations in $\mathcal{U}_{\text{LJ}}\left(\textbf{r}^n\right)$ are suppressed (Fig.~\ref{fig:Fcc-int}), and the system effectively behaves as a non-interacting Einstein crystal. In this limit, the Helmholtz free energy can be estimated analytically as:
\begin{eqnarray}
f_x(T,\gamma_{\max}) &=& \langle \mathcal{U}_{\text{LJ}}\rangle_{\gamma_{\max}} - \frac32\,kT\ln\frac{2\pi kT}{\gamma_{\max}},
\end{eqnarray}
where $\langle \mathcal{U}_{\text{LJ}}\rangle_{\gamma}$ is the average potential energy at a given  $\gamma$. The Helmholtz free energy at $\gamma=0$ is then obtained via thermodynamic integration:
\begin{eqnarray}
f_x(T;0) &=& \frac{F_x(T;0)}{n} \notag\\
 &=& f_x(T;\gamma_{\max}) - \frac{1}{2\sigma_{AA}^2}\int_0^{\gamma_{\max}}\left\langle|\textbf{r}-\textbf{r}_0|^2 \right\rangle_{\gamma}\,d\gamma\notag\\
 && \label{eq:FL-integral}
\end{eqnarray}
The corresponding chemical potential is given by  $\mu_x = f_x + p\langle v_x\rangle_p$, where $\langle v_x\rangle_p$ is the specific volume of the crystal at pressure $p$. 

To perform the integration in Eq.~\eqref{eq:FL-integral}, we first conduct $NpT$  simulations of the \textsc{Fcc} crystal with $n=13,500$ particles to determine its zero-pressure density, which is found to be $\rho^*=\rho\sigma_{AA}^3=0.9776$ at $kT/\epsilon_{AA}=0.5$. Canonical ensemble simulations are then carried out at this density to evaluate the mean squared displacement across a wide range of $\gamma$ values. In total, 201 simulations are performed wherein $\gamma$ is logarithmically varied between $10^{-3}$ and $4.85\times10^5$. The largest $\gamma$ is chosen so that $\langle\mathcal{U}_{\text{LJ}}\rangle_{\gamma}$ plateaus to the ideal lattice value, and the per-particle mean squared displacement approaches the canonical behavior of a harmonic oscillator (Fig.~\ref{fig:Fcc-int}), namely:
\begin{eqnarray}
\langle |\textbf{r}-\textbf{r}_0|^2\rangle_\gamma = \frac{3kT}{2\gamma}.\label{eq:limiting-harmonic-MSD}
\end{eqnarray}
To avoid known numerical instabilities associated with the Nos\'{e}-Hoover thermostat when applied to stiff oscillators,\cite{MartynaJCP1992} we use a Langevin thermostat with a time step of $10^{-4}$ in reduced \textsc{Lj} units.

Once the chemical potentials of both the crystal and supercooled liquid are computed at  $kT/\epsilon_{AA}=0.5$ and $p=0$, we extend conventional equations of state calculation to neighboring state points. In particular, we conduct additional isothermal isobaric simulations to estimate $\mu$ over the ranges $kT/\epsilon_{AA}\in [0.5,0.67]$ and $p\sigma_{AA}^3/\epsilon_{AA}\in [0,0.05]$, enabling us to evaluate the chemical potential as a function of temperature and pressure for both phases.

\subsection{Coexistence simulations}
\label{section:coexistence}

\noindent
As an additional consistency check, we perform coexistence simulations to determine the melting temperature at $p^*=0.05$. Initial configurations are generated by initiating short (50,000-step) \textsc{Md} trajectories from a cuboidal box containing 12,960 \textsc{Lj} particles arranged in an \textsc{Fcc} lattice, with a reduced number density of $\rho_n^*=1.13$. During these simulations, which are conducted in the canonical ensemble at $kT/\epsilon_{AA}=2$, only particles within one half of the box are allowed to move, causing the crystal in the corresponding region to melt. The endpoints of these simulations are then quickly relaxed in the isothermal-isobaric ensemble at the designated temperature and pressure. At each temperature, 10 independent configurations are generated.

\section{Results and Discussions}

\subsection{Sensitivity of nucleation kinetics to temperature}

\begin{table}
    \centering
    \caption{Homogeneous nucleation rates in the \textsc{Lj} system at $p=0$. Data taken from Refs.~\citenum{Haji-Akbari2018Forward-fluxParameters} and \citenum{HussainJCP2022}. Reported uncertainties correspond to 95\% confidence intervals.}
    \label{table:LJ_homrates}
    \begin{tabular}{c|c} 
        \hline\hline
        $kT/\epsilon_{AA}$  & $\log_{10}R_{\text{hom}} \left[\sigma_{AA}^{-4}\epsilon_{AA}^{-1/2}m_A^{-1/2}\right]$\\ [0.5ex] 
        \hline
        0.44 & ~$-6.9931 \pm 0.0810$ \\ 
        0.45 & ~$-7.8193 \pm 0.0930$ \\
        0.46 & ~$-9.3918 \pm 0.0986$\\
        0.47 & $-11.1376 \pm 0.0927$\\ 
        0.48 & $-13.3593 \pm 0.0998$\\
        0.50 & $-18.6180 \pm 0.0840$\\ [1ex] 
        \hline\hline
    \end{tabular}
\end{table}

\noindent
In order to probe the susceptibility of \textsc{Cnt} to chemical heterogeneity of a crystal nucleating surface, we first examine its ability to describe the temperature dependence of nucleation kinetics under standard conditions. Firstly, we consider homogeneous nucleation. Assuming that fusion enthalpies and entropies are weak functions of temperature, $R_{\text{hom}}$, the homogeneous nucleation rate, will exhibit the following dependence on temperature:\cite{Cabriolu2015IceNucleation}
\begin{eqnarray}\label{eq:CNT-T-hom}
\ln R_{\text{hom}} &=& \ln A_{\text{hom}} + \frac{C_{\text{hom}}}{T(T-T_m)^2},
\end{eqnarray}
with $T_m$, the equilibrium melting temperature.  Fig.~\ref{fig:CNT_fits}c depicts that Eq.~(\ref{eq:CNT-T-hom}) successfully describes the temperature dependence of the nucleation rates reported in Refs.~\citenum{Haji-Akbari2018Forward-fluxParameters} and \citenum{HussainJCP2022} and given in Table~\ref{table:LJ_homrates}.

We then consider heterogeneous nucleation on a flat uniform surface comprised of \textsc{Lj} particles that interact with liquid particles via $\epsilon_{AB}=0.5\epsilon_{AA}$. and $\sigma_{AB}=\sigma_{AA}$. Chemical uniformity implies the feasibility of maintaining a fixed contact angle, as assumed by \textsc{Cnt}, which, upon combining Eqs.~(\ref{eq:DelG-het}) and (\ref{eq:CNT-T-hom}), will yield the following scaling of $R_{\text{het}}$ with $T$:\cite{Cabriolu2015IceNucleation}
\begin{eqnarray}\label{eq:CNT-T-het}
\ln R_{\text{het}} &=& \ln A_{\text{het}} + \frac{C_{\text{het}}}{T(T-T_m)^2},
\end{eqnarray}
where $C_{\text{het}}=f_c(\theta_c)C_{\text{hom}}$. 
The computed heterogeneous nucleation rates (Table~\ref{table:LJ_hetrates_good}) are also depicted in Fig.~\ref{fig:CNT_fits}c and are satisfactorily described by Eq.~(\ref{eq:CNT-T-het}). 
 These observations confirm the baseline ability of \textsc{Cnt} to properly describe the canonical dependence of nucleation kinetics on $T$ in the \textsc{Lj} system in both homogeneous and heterogeneous nucleation. 
 

\begin{figure}
    \centering
    \includegraphics[width=.4\textwidth]{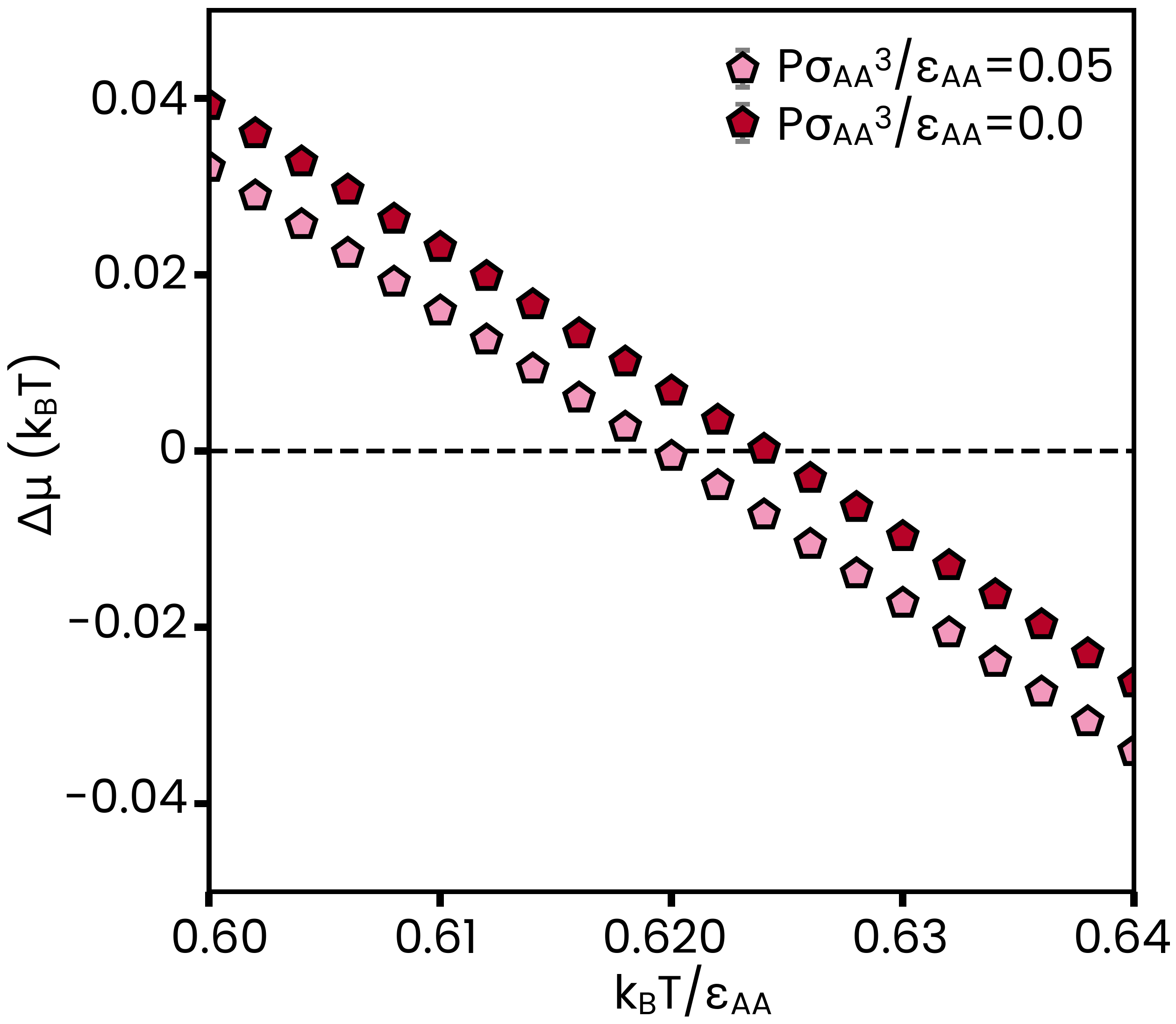}
    \caption{\label{fig:freenergycalc} Chemical potential difference between the liquid and the face-centered cubic (\textsc{Fcc}) crystal as a function of temperature at two different pressures. Error bars are smaller than symbol sizes.}
\end{figure}

We then explore crystal nucleation on a checkerboard surface comprised of alternating liquiphilic and liquiphobic patches. Each patch is almost square-shaped with a side length of $\approx9\sigma_{AA}$. Particles in liquiphilic patches are identical to those in the uniform surface considered above, while liquiphobic particles interact with liquid particles via the \textsc{Wca} potential\cite{WeeksJChemPhys1971} with $\epsilon_{AC}=0.3\epsilon_{AA}$. Intuitively, nucleation will be initiated on liquiphilic patches, but the growing nuclei will be forced to adopt geometries that could make the sustenance of a fixed contact angle impossible. Nonetheless, heterogeneous nucleation rates (Table~\ref{table:LJ_hetrates_chck}) readily follow Eq.~\eqref{eq:CNT-T-het} as depicted in Fig.~\ref{fig:CNT_fits}c, demonstrating the surprising robustness of \textsc{Cnt} to regular nanoscale chemical heterogeneity on crystal nucleating surfaces.

\subsection{Changes in melting temperature}

\noindent
Despite the surprisingly superb performance of \textsc{Cnt}, we observe an intriguing anomaly in our \textsc{Cnt} fits. When treating $T_m$  as a fitting parameter in Eqs.~\eqref{eq:CNT-T-hom} and \eqref{eq:CNT-T-het} (dashed lines in Fig.~\ref{fig:CNT_fits}c), we obtain $kT_m/\epsilon_{AA}=0.635\pm0.020$ for homogeneous nucleation and $kT_m/\epsilon_{AA}=0.609\pm0.010$ for heterogeneous nucleation on the checkerboard surface. (A reliable estimate of $T_m$ for the uniform surface is not possible due to an insufficient number of rate calculations.) Not only do these fitted values differ across different modes of nucleation, but they are also significantly lower than the widely reported melting temperature of the \textsc{Lj} system at zero pressure-- $kT_m/\epsilon_{AA}\approx0.69$.\cite{Agrawal1995ThermodynamicCoexistenceb}

To reconcile this discrepancy, we conduct free energy calculations using a modified\cite{Haji-Akbari2011PhaseTetrahedra} variant of the Frenkel-Ladd approach\cite{FrenkelJCP1984} discussed in Section~\ref{section:free-energy-calcs}. Fig.~\ref{fig:freenergycalc} shows the chemical potential difference, $\Delta\mu=\mu_{\text{liq}}-\mu_{\text{sol}}$, as a function of temperature for $p\sigma_{AA}^3/\epsilon_{AA}=0.0$ and $0.05$, yielding melting temperatures of $kT_m/\epsilon_{AA}=0.619$ and $0.624$, respectively.

\begin{figure*}
    \centering
    \includegraphics[width=.7\textwidth]{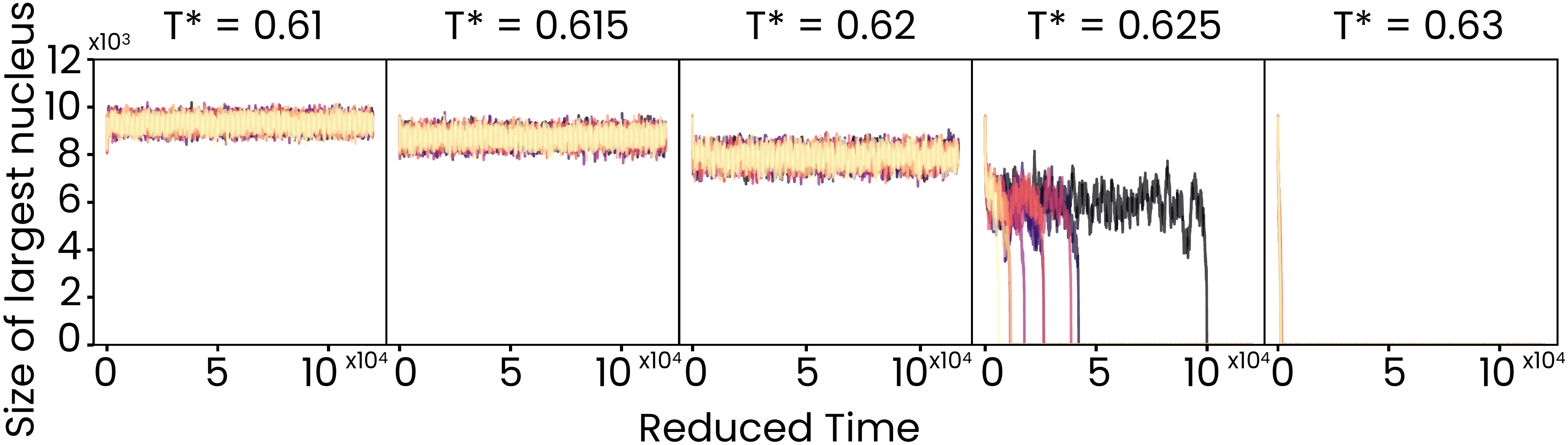}
    \caption{\label{fig:coexistence_plot} Temporal evolution of the size of the largest crystalline nucleus in coexistence simulations at $p\sigma_{AA}^3/\epsilon_{AA}=0.05$ at different temperatures. The system melts completely for simulations at or above $kT/\epsilon_{AA}\approx0.625$. Different colors correspond to trajectories initiated from different starting configurations.}
\end{figure*}

We further corroborate these findings using coexistence simulations using the procedure discussed in Section~\ref{section:coexistence} at temperatures around  $kT_m/\epsilon_{AA}=0.624$. 
Each coexistence simulation is performed for a minimum of $4\times10^7$ steps. As illustrated in Fig.~\ref{fig:coexistence_plot}, we observe a sharp decline in the size of the largest crystalline nucleus for temperatures,  $kT/\epsilon_{AA} > 0.62$, culminating in the complete melting of the crystal. Conversely, for systems at $kT/\epsilon_{AA}\leq 0.62$, there is no considerable change in the size of the largest nucleus throughout our extended simulations, even when extended beyond the duration illustrated in Fig.~\ref{fig:coexistence_plot}. Note that coexistence simulations are always fairly slow at fully melting and/or crystallizing when the target temperature is very close to the equilibrium melting temperature. This temperature range coincides with the melting temperature estimated from free energy calculations and the adjustable fitting parameter from the \textsc{Cnt} fit.

\begin{figure}
 \centering
 \includegraphics[width=0.5\textwidth]{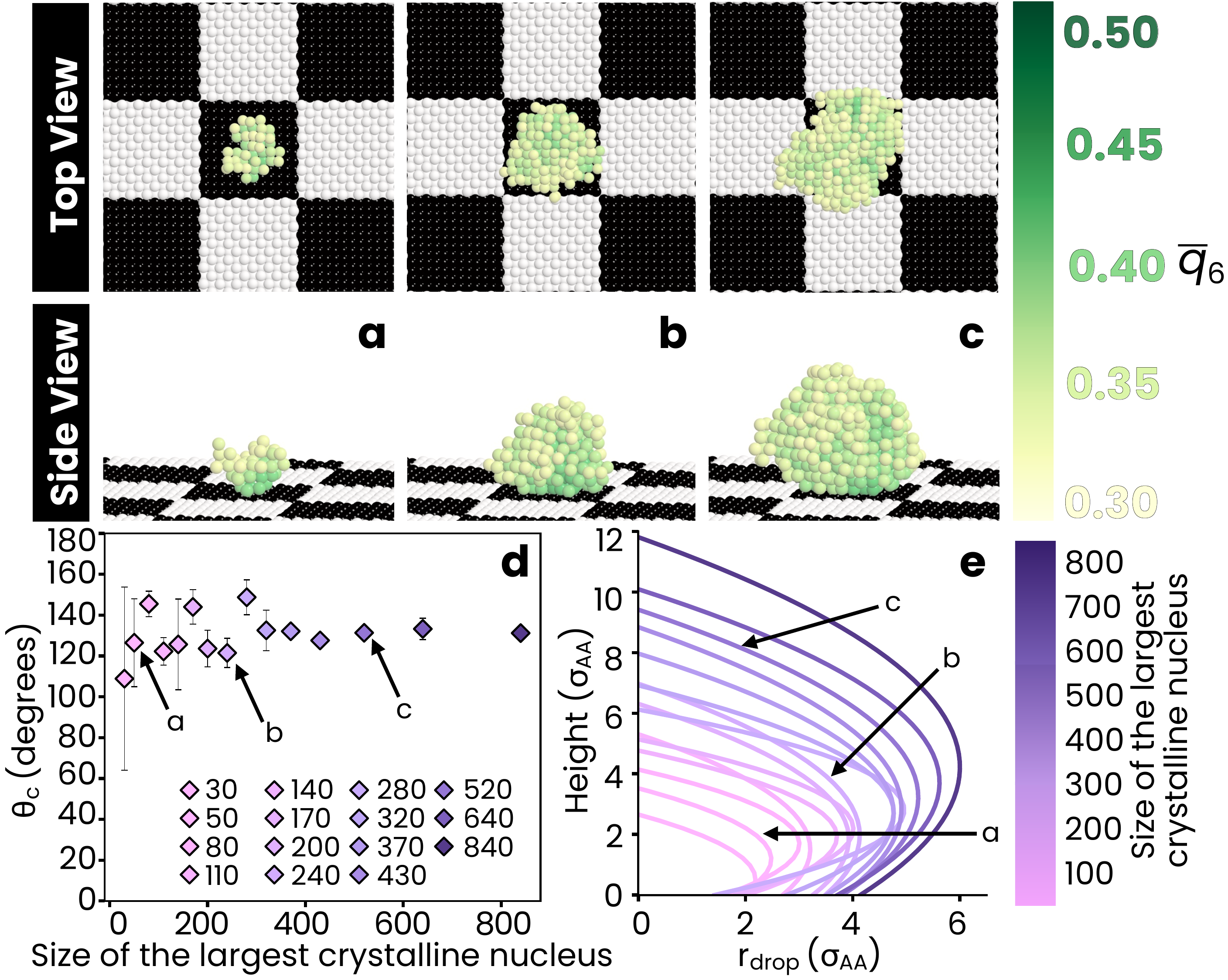}
 \caption{\label{fig:clusters} (a-c) Temporal evolution of the largest crystalline nucleus in the checkerboard system at $k_BT/\varepsilon_{AA}=0.536$. (a) A crystalline nucleus forms within a liquiphilic patch, and (b) approaches the patch boundary upon further growth, (c) resulting in vertical growth towards the bulk. Particles are colored based on their local neighbor-averaged Steinhardt bond order parameter, $\overline{q}_6$, values. (d) Contact angle vs.~nucleus size for all \textsc{jFfs} milestones at  $k_BT/\varepsilon_{AA}=0.536$. (e) $r_{\text{drop}}(z)$ for nuclei of different sizes, highlighting how the nucleus shape evolves during the nucleation process.
 }
\end{figure}

These findings highlight the sensitivity of melting thermodynamics of the truncated and shifted \textsc{Lj} potential to the choice of cutoff radius. (We wish to note that the $T_m$ reported in Ref.~\citenum{Agrawal1995ThermodynamicCoexistenceb} is for the truncated potential with a long-range tail correction.) A detailed investigation into how the cutoff radius and shifting impacts nucleation kinetics is beyond the scope of this work and will be addressed in future studies.

\begin{table*}
\centering
\caption{Fitting parameters, and the corresponding contact angles and potency factors obtained from Eqs.~\eqref{eq:CNT-T-hom} and \eqref{eq:CNT-T-het} by using the $T_m$ values estimated from direct free energy calculations.\label{tab:contact-angles}}
\begin{tabular}{|l|cc|c|c|}
\hline
System & $A$ & $C$ & $f_c(\theta_c)$ & $\theta_c$ \\
\hline
Homogeneous & $5.136\pm0.330$ & $-0.1708\pm0.0034$ & -- & -- \\
Uniform & $11.026\pm3.025$ & $-0.0652\pm0.0085$ & $0.3816\pm0.0502$ & $~80.84^{\circ}\pm3.99^{\circ}$ \\
Checkerboard & $14.358\pm0.402$ & $-0.1567\pm0.0023$ & $0.9172\pm0.0226$ & $130.25^{\circ}\pm3.88^{\circ}$\\
\hline
\end{tabular}
\end{table*}

\subsection{Microscopic validity of classical nucleation theory}

\noindent
We then examine the microscopic validity of a core assumption of \textsc{Cnt}-- that the contact angle remains constant throughout the nucleation process-- by first visually observing the geometric evolution of crystalline nuclei.  Intuitively, crystalline nuclei are expected to form within liquiphilic patches and can grow at a fixed contact angle while being contained within the starting patch. However, as they grow laterally, they will encounter adjacent liquiphobic regions that will inhibit further spreading and disrupt maintaining the ideal spherical cap geometry assumed in \textsc{Cnt}. Continued growth may proceed either by 'bridging` into neighboring liquiphilic patches or by expanding vertically into the bulk liquid.

 Figs.~\ref{fig:clusters}a-c illustrate the geometric evolution of a representative crystalline nucleus.  As expected, nucleation initiates within a liquiphilic patch (Fig.~\ref{fig:clusters}a), and the growing nucleus avoids extending into adjacent liquiphobic regions (Fig.~\ref{fig:clusters}b). Rather than bridging laterally, the nucleus predominantly grows upward, away from the surface, and into the supercooled liquid (Fig.~\ref{fig:clusters}c). 

To quantitatively characterize the evolution of nucleus shapes during nucleation, we calculate the average contact angles of nuclei collected at various \textsc{jFfs} milestones, using the methodology discussed in Section~\ref{section:contact-angle-calc}. As depicted in Fig.~\ref{fig:clusters}d, the contact angles show negligible dependence on nucleus size. Except for some earlier milestones, where the computed contact angles exhibit greater uncertainties due to the smaller sizes of the corresponding nuclei, the remaining contact angles  consistently hover  between 120$^\circ$ and 140$^\circ$. Therefore, despite considerable chemical heterogeneity, the effective contact angle remains virtually unchanged. 

A more interesting aspect of the nucleation mechanism is revealed upon examining the $r_{\text{drop}}$ profiles used for computing the contact angles. These profiles, presented in Fig.~\ref{fig:clusters}e, quantitatively encode the shapes of the crystalline nuclei and suggest a consistent nucleus geometry across different nucleus sizes. Remarkably, larger nuclei tend to become pinned at approximately $4\sigma_{AA}$, which is roughly half the size of the liquiphilic patch. Therefore, nuclei that form at the center of a liquiphilic patch become pinned at the patch boundaries, as depicted in Figs.~\ref{fig:clusters}b-c. Furthermore, the heights of the nuclei increase continuously with size, corroborating the vertical growth scenario observed visually in Fig.~\ref{fig:clusters}c. This boundary pinning and vertical growth, in turn, help nuclei preserve their contact angle.

\begin{figure}
    \centering
    \includegraphics[width=.45\textwidth]{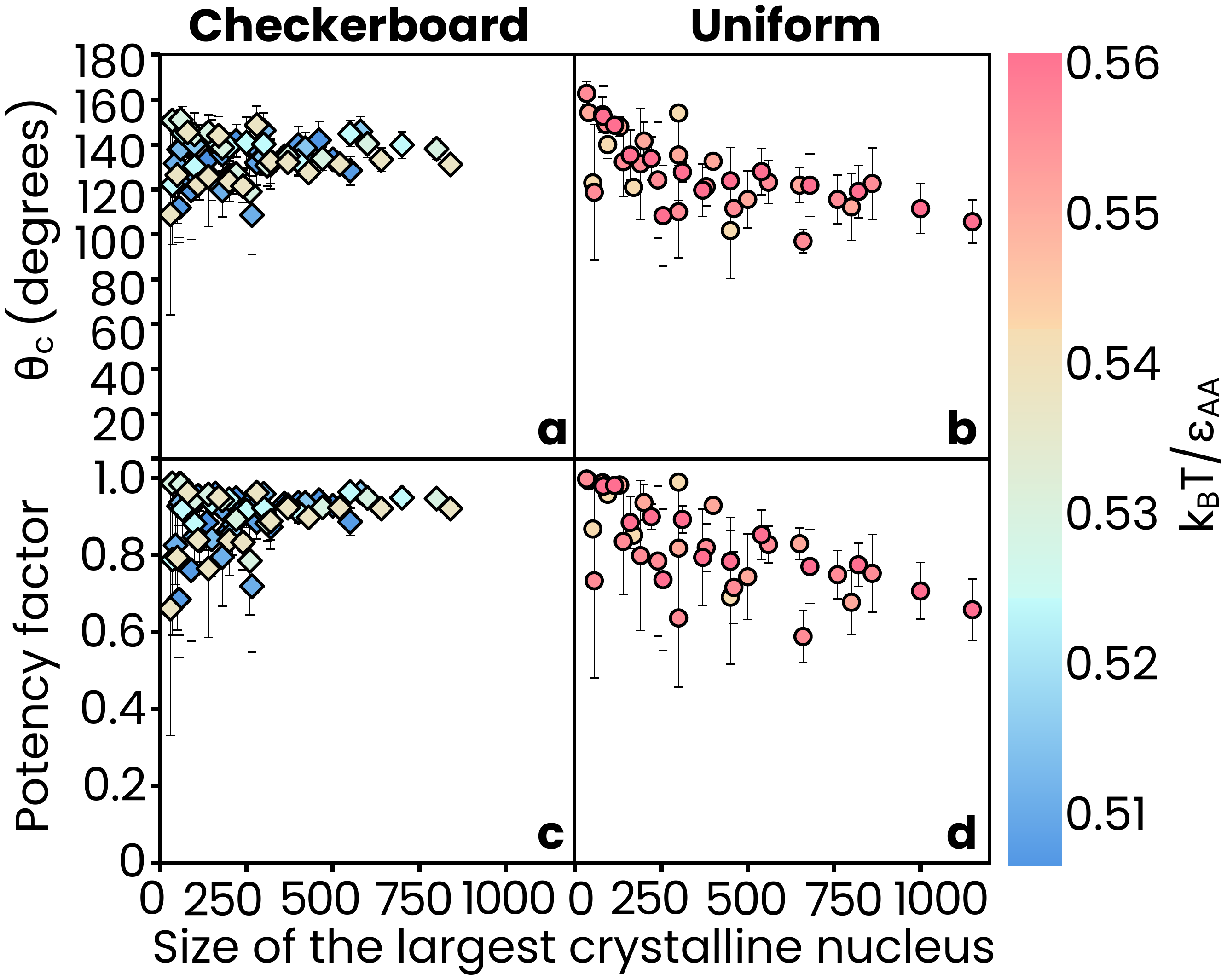}
    \caption{\label{fig:contactangle} Geometric estimates of (a-b) contact angles, $\theta_c$, and (c-d) potency factors,  $f_c(\theta_c)$, of crystalline nuclei in the vicinity of (a,c) checkerboard and (b,d) uniform surfaces. 
    }
\end{figure}

\begin{figure}
    \centering
    \includegraphics[width=.45\textwidth]{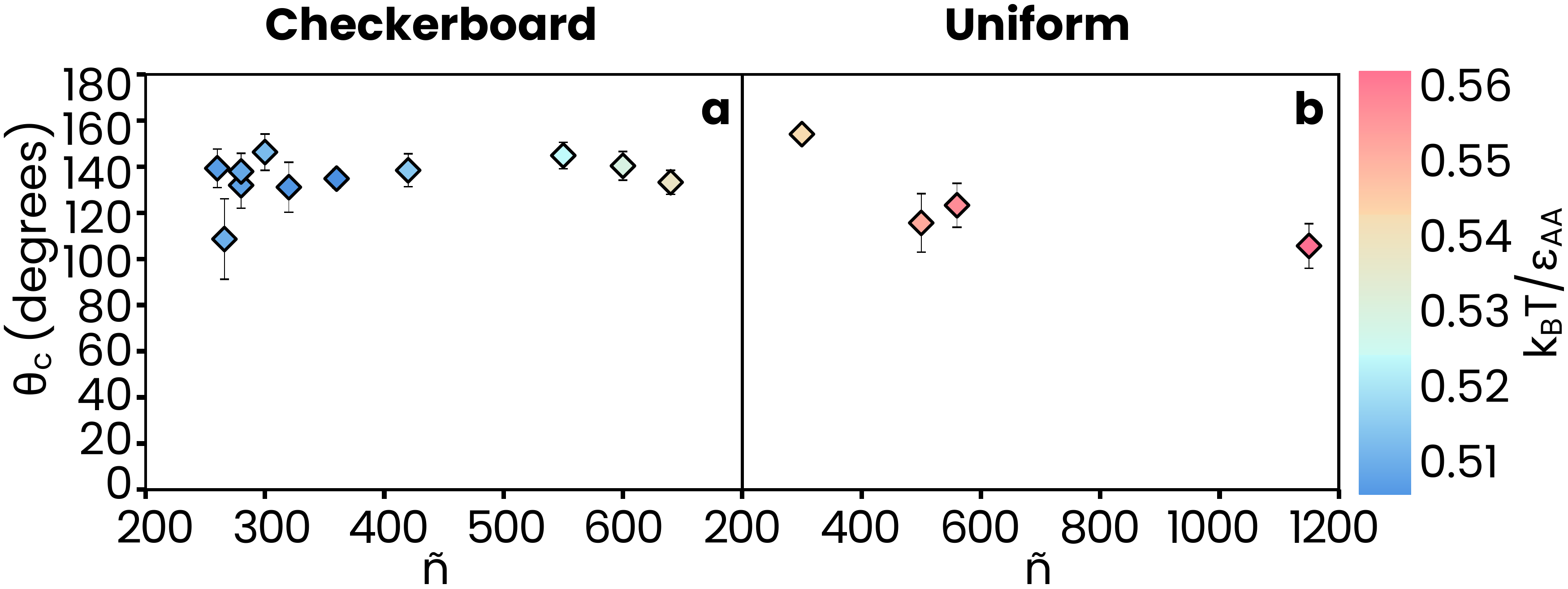}
    \caption{\label{fig:contactangle-critical} Geometric estimates of the average contact angle of a critical nucleus at different temperatures in (a) the checkerboard and (b) the uniform system. $\tilde{n}$ refers to the \textsc{jFfs} milestone closest to $n^*$, the critical nucleus size at a given temperature, determined from fitting an error function to survival probabilities.}
\end{figure}

These observations hold upon examining contact angles and $r_{\text{drop}}$ profiles at other temperatures. As illustrated in Figs.~\ref{fig:contactangle}a and \ref{fig:contactangle}c, the computed average contact angles and potency factors across all temperatures and \textsc{jFfs} milestones exhibit minimal variability. This finding explains the efficacy of \textsc{Cnt} in accurately describing the canonical temperature dependence of nucleation kinetics within the checkerboard system. Additionally, the contact angles for critical nuclei-- defined as those with an approximately 50\% survival probability as per Ref.~\citenum{Domingues2024DivergenceCrystals}-- remain relatively stable across the examined temperature range (Fig.~\ref{fig:contactangle-critical}a).

We also compute contact angles (Fig.~\ref{fig:contactangle}b) and potency factors (Fig.~\ref{fig:contactangle}d) for crystalline nuclei emerging on the uniform surface.  Surprisingly, the contact angles exhibit slightly larger variability in the chemically uniform system and tend to decrease with increasing nucleus size, evidenced by a weighted linear regression $R^2$ value of 0.65. (A similar weak decline is observed upon examining contact angles of critical nuclei, as illustrated in Fig.~\ref{fig:contactangle-critical}b.)  This is in contrast to the checkerboard system, wherein contact angles exhibit a weaker size dependence with a smaller linear slope and an $R^2$ value of 0.41 (Fig.~\ref{fig:contactangle}a). This counterintuitive observation likely arises from the distinct growth mechanisms in these two systems. As discussed above, nuclei on the checkerboard surface evolve in accordance with a pinning mechanism, preferentially avoiding liquiphobic patches and growing vertically into the bulk.  In contrast, nuclei on the uniform surface can freely adopt a wider range of shapes, governed by a complex interplay between liquid–wall interactions and the anisotropic surface energies of different crystalline facets.

We then compare the potency factors obtained from the geometric estimates of contact angles with those inferred from \textsc{Cnt}. For the latter, rather than treating $T_m$ as a fitting parameter,  we use values computed from free energy calculations at $p=0$ for homogeneous nucleation and at  $p\sigma_{AA}^3/\epsilon_{AA}=0.05$ for heterogeneous nucleation. $f_c$ is then estimated as $f_c=C_{\text{het}}/C_{\text{hom}}$ for every surface, following the approach of Ref.~\citenum{Cabriolu2015IceNucleation}.  The resulting values are summarized in Table~\ref{tab:contact-angles}. For the checkerboard surface, the \textsc{Cnt}-inferred potency factor of $0.917\pm0.023$ is fairly consistent with the geometric estimate, $\approx0.901\pm0.066$.  In contrast, for the chemically uniform surface, the \textsc{Cnt}-inferred value, $f_c=0.382\pm0.050$, is substantially smaller than the average geometric estimate, $0.830\pm0.160$. 

This substantial discrepancy can be attributed to several factors. Most importantly, the specifics of the utilized order parameter can significantly influence the sizes and shapes of crystalline nuclei. In contrast, nucleation rates,  which constitute the bases for \textsc{Cnt} estimates of contact angle, tend to be less sensitive to such details.\cite{Hussain2020StudyingOutlook, Domingues2024DivergenceCrystals} The order parameter employed in this work was specifically optimized in our earlier work for probing nucleation on the checkerboard surface,\cite{Domingues2024DivergenceCrystals} while its efficacy in the case of the chemically uniform surface is not fully established.   Moreover, given the stronger dependence of $\theta_c$ on nucleus size in the chemically uniform surface, the pertinent range of nucleus sizes relevant for potency factor calculations remains unclear.  In addition to these complications, it is important to note that the mean-field concept of a contact angle in \textsc{Cnt} does not necessarily have to translate to geometric estimates. Therefore, the agreement between the two for the checkerboard surface can simply be coincidental. 

\begin{figure*}
    \centering
    \includegraphics[width=.9\textwidth]{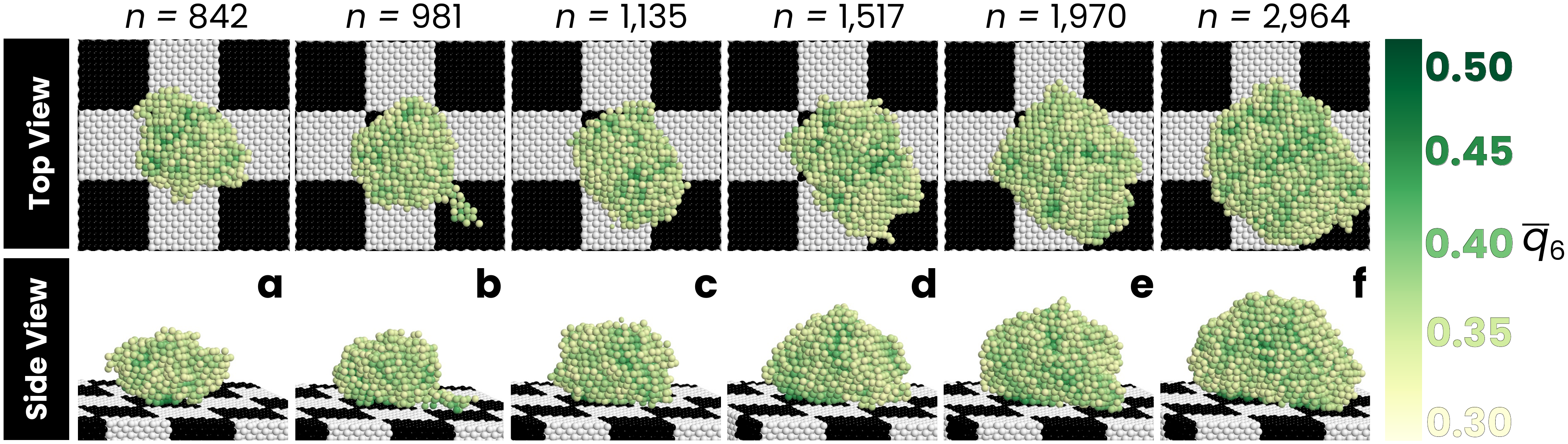}
    \caption{\label{fig:pinning} Growth of a post-critical nucleus on the checkerboard surface at $k_BT/\epsilon_{AA}=0.536$. The initial nucleus is comprised of 842 particles, and the subsequent nuclei are captured at 12,000-step intervals. Note the prevalence of pinning at the patch boundaries. 
    }
\end{figure*}

\begin{figure}
    \centering
    \includegraphics[width=.4\textwidth]{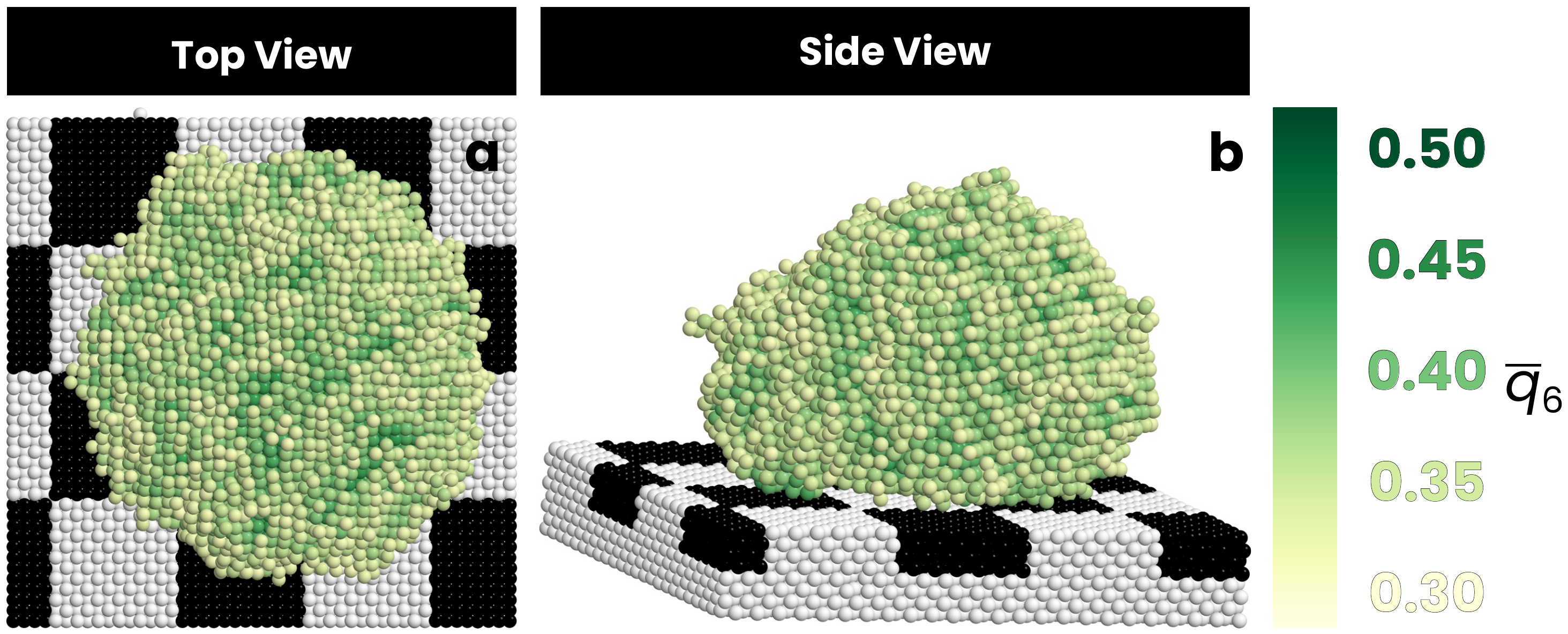}
    \caption{\label{fig:growth} A large post-critical crystalline nucleus comprised of 8,337 particles at $k_BT/\epsilon_{AA}=0.536$, establishing multiple lateral bridges into neighboring liquiphilic domains while still avoiding the patch boundaries. 
    }
\end{figure}

\subsection{Boundary pinning during crystal growth}

\noindent
To investigate the relevance of boundary pinning to crystal growth, we launch conventional \textsc{Md} simulations from several post-critical configurations at $k_BT=0.536$. (Those configurations are chosen from the crossing events at the last \textsc{jFfs} milestone.)  Figure~\ref{fig:pinning} displays six snapshots from one such trajectory, which originates from a post-critical configuration that is a progeny of the nucleus depicted in  Fig.~\ref{fig:clusters}c. These snapshots, which are separated by 12,000 steps, demonstrate that the growing nuclei systematically avoid the liquiphobic patches, as is evident from the  side view panel of Fig.~\ref{fig:pinning}. The growth mechanism thus involves a combination of vertical expansion and diagonal bridging into adjacent liquiphilic patches. Notably, these diagonal bridges frequently break (Fig.~\ref{fig:pinning}c) and reform (Fig.~\ref{fig:pinning}d) during the growth process. It is noteworthy that diagonal bridge formation is not observed during nucleation at the considered temperature range. These observations can be attributed to the weak energetic attractions between the particles within the liquid and those in the liquiphilic patches, making the adoption of solid-like structures at the bridge points less energetically favorable (than in the bulk).


Figure~\ref{fig:growth} presents a crystalline nucleus at a more advanced stage of growth, wherein multiple diagonal bridges have already formed. Despite the emergence of these bridges, the nucleus continues to avoid direct contact with liquiphobic patches, maintaining pinning at the patch boundaries. As crystallization progresses, the nuclei are expected to eventually engulf the liquiphobic patches, yet continue to circumvent them, resulting in a corrugated liquid/solid interface along the surface of the crystal.

\section{Conclusions}

\noindent
In this work, we employ \textsc{Md} simulations, \textsc{jFfs}, and free energy calculations to examine the temperature dependence of heterogeneous crystal nucleation rates within a model \textsc{Lj} liquid on a weakly attractive chemically uniform surface and a checkerboard surface comprised of liquiphilic and liquiphobic patches. Our findings reveal that \textsc{Cnt} provides a reasonable description of the temperature dependence of computed rates on both surfaces. By directly evaluating the contact angles of individual nuclei at successive stages of nucleation, we find that the contact angle exhibits far less variability and remains nearly constant on the patterned substrate, despite its chemical heterogeneity. This behavior arises from the vertical growth of nuclei that become pinned at patch boundaries, thereby preserving a stable apparent contact angle. Interestingly, contact angles are slightly more variable in the chemically uniform system. However, such variability is too small to translate into a  breakdown in \textsc{Cnt}'s predictive power.

These findings highlight the complex relationship between the applicability of \textsc{Cnt} and surface chemistry. Yet, they still offer insights into why \textsc{Cnt} has proven remarkably effective, even when the underlying crystal nucleating surfaces possess complex chemistries. Indeed, chemically and topographically heterogeneous surfaces often feature a mosaic of highly active nucleating regions interspersed with inert domains. Such spatial variability can give rise to a pinning mechanism in which crystalline nuclei grow vertically out of active patches, and maintain relatively constant contact angles along the edges of active-passive domains, thereby satisfying the key assumption of \textsc{Cnt}. If a non-uniform surface is comprised of patches of varying (but positive) nucleating tendencies, pinning of the type discussed here will no longer be feasible, potentially resulting in major deviations from \textsc{Cnt}. However, such cases appear to be the exception rather than the rule.

It is important to highlight that our metric for the performance of \textsc{Cnt} is  deliberately minimalistic. Specifically, we focus solely on the correct functional dependence of the nucleation rate on temperature, without demanding quantitative accuracy for kinetic prefactors and the nucleation barrier, as well as other key assumptions of \textsc{Cnt}. Indeed, earlier studies have found  \textsc{Cnt}  to fail when scrutinized at such a granular level.\cite{CacciutoPRL2004, BaidakovJPCB2019} However, we argue that our performance metric is more pertinent to experiments, where measured rates are typically fitted to the functional form predicted by \textsc{Cnt} rather than attempting to estimate individual parameters independently. 

Given the complex interplay between surface heterogeneity and the validity of \textsc{Cnt}, further investigations on patterned surfaces are essential to pinpoint the specific combinations of geometries and chemistries that could lead to the breakdown of CNT. Although such potential breakdowns may complicate the accurate modeling of such surfaces, they also offer an opportunity to design crystal nucleating surfaces that enable enhanced control over crystal texture and grain size distribution.



\begin{acknowledgments}
\noindent 
A.H.-A. gratefully acknowledges the support of National Science Foundation (\textsc{Nsf}) Grants \textsc{Cbet}-1751971 (\textsc{Career} Award). This work was also supported by the Alfred P. Sloan Foundation.  
F.S.V. acknowledges financial support by the \textsc{Gem} fellowship and the National Science Foundation Graduate Research Fellowships Program (\textsc{Nsf-Grfp}).
These calculations were performed at the Yale Center for Research Computing.  This work used the Extreme Science and Engineering Discovery Environment (\textsc{Xsede}), which is supported by \textsc{Nsf} Grant \textsc{Aci}-1548562. This work used Stampede through allocation \textsc{Chm}240063 from the Advanced Cyberinfrastructure Coordination Ecosystem: Services \& Support (\textsc{Access}) program, which is supported by \textsc{Nsf} Grants \#2138259, \#2138286, \#2138307, \#2137603, and \#2138296.
\end{acknowledgments}


\bibliographystyle{achemso}
\bibliography{references}

\end{document}